\documentclass[onecolumn,showpacs,preprintnumbers]{revtex4}
\usepackage{mathrsfs}
\usepackage{graphicx}
\usepackage{dcolumn}
\usepackage{bm}
\usepackage{epsfig}
\usepackage{amsmath,amssymb}

\begin{document}
\title{The ground states and the first radially excited states of D-wave vector $\rho$ and $\phi$ mesons}
\author{Guo-Liang Yu$^{1}$}
\email{yuguoliang2011@163.com}
\author{Zhi-Gang Wang$^{1}$}
\email{zgwang@aliyun.com}
\author{Xiu-Wu Wang$^{1}$}
\author{Hui-Juan Wang$^{1}$}

\affiliation{Department of Mathematics and Physics, North China
Electric power university, Baoding 071003, People's Republic of
China}
\date{\today }

\begin{abstract}
In this article, we firstly derive two QCD sum rules QCDSR I and QCDSR II which are respectively used to extract observable quantities of the ground states and the first radially excited states of D-wave vector $\rho$ and $\phi$ mesons. In our calculations, we consider the contributions of vacuum condensates up to dimension-7 in the operator product expansion. The predicted masses for $1^{3}D_{1}$ $\rho$ meson and $2^{3}D_{1}$ $\phi$ meson are consistent well with the experimental data of $\rho$($1700$) and $\phi$($2170$). Besides, our analysis indicates that it is reliable to assign the recent reported $Y$($2040$) state as the $2^{3}D_{1}$ $\rho$ meson. Finally, we obtain the decay constants of these states with QCDSR I and QCDSR II.
These predictions are helpful not only to reveal the structure of the newly observed $Y$($2040$) state but also to establish $\phi$ meson and $\rho$ meson families.\end{abstract}


\maketitle

\begin{large}
\textbf{1 Introduction}
\end{large}

Recently, BESIII Collaboration reported two resonant structures in measuring born cross sections for the processes $e^{+}e^{-}\rightarrow\omega\eta$ and $e^{+}e^{-}\rightarrow\omega\pi^{0}$ for center-of-mass energies between $2.00$ and $3.08$ GeV\cite{BESIII}.
In the process $e^{+}e^{-}\rightarrow\omega\eta$, a resonance with a mass of ($2179\pm21\pm3$)MeV$/c^{2}$ and a width of ($89\pm28\pm5$)MeV was observed with a significance of $6.1\sigma$. This structure is consistent with the properties of $\phi(2170)$ resonance. Another structure with a mass of ($2034\pm14\pm9$)MeV$/c^{2}$, width of ($234\pm11\pm$)MeV was observed in the $\omega\pi^{0}$ cross section. By analyzing its mass and decay width, BESIII Collaboration assigned this structure to be the $\rho(2000)$ state, which had been suggested to be a $2^{3}D_{1}$ $\rho$ meson by Regge trajectory analysis\cite{rou21,LPHe}.

$\rho(2000)$ was first observed in the $p\overline{p}\rightarrow\pi\pi$ process with a mass of $1988$ MeV\cite{Hasan}. Later, its existence was also confirmed in the processes $p\overline{p}\rightarrow\omega\eta\pi^{0}$ and $p\overline{p}\rightarrow\omega\pi$\cite{rou21,rou22,rou23,rou24}. By analyzing its decay properties and Regge trajectory, scientists suggested $\rho(2000)$ was the first radial excitation of $\rho(1700)$\cite{rou21,LPHe}. However, $\rho(2150)$ was also predicted to be the $2^{3}D_{1}$ state with $2.15$ GeV by Godfrey and Isgur\cite{GI}, which is contradictory with this above conclusion about $\rho(2000)$. In ref.\cite{Bugg1}, Bugg speculated $\rho(2000)$ to be a mixed state with a significant component of $^{3}D_{1}$. In Ref.\cite{Ebert}, the masses of $\rho(1700)$ and $\rho(2000)$ were predicted to be about $1.570$GeV and $1.900$GeV\cite{BESIII} which are lower than the experimental data by $\sim$ $100$ MeV. In order to clarify all of these questions, it is necessary to make a theoretical analysis about the ground state and the first radially excited state of $\rho$ meson.

$\phi(2170)$ state was also denoted as $X(2170)$ or $Y(2170)$ in some literatures and it was observed by the BARBAR Collaboration in the process $e^{+}e^{-}\rightarrow\gamma\phi f_{0}$\cite{Aubert}. Later, its existence was confirmed by Belle\cite{CPShen}, BESII\cite{Ablikim1} and BESIII\cite{Ablikim2,Ablikim3,Ablikim4,Ablikim5,Ablikim6} Collaborations. Since the observation of $\phi(2170)$, scientists have made considerable efforts to understand its natrue and have proposed many interpretations such as a conventional $3^{3}S_{1}$ or $2^{3}D_{1}$ $s\overline{s}$ state\cite{MGI,GJDing,XWang,CGZhao,QLi,Coito,Badalian,Barnes}, an $s\overline{s}g$ hybrid\cite{phissg1,phissg2}, a tetraquark state\cite{phi4s1,phi4s2,phi4s3,phi4s4,phi4s5,phi4s6,phi4s7,phi4s8,phi4s9,phi4s10}, a $\Lambda\overline{\Lambda}$ bound state\cite{phi2Lamd1,phi2Lamd2,phi2Lamd3,phi2Lamd4}, as well as $\phi K\overline{K}$\cite{phikk1,phikk2} and $\phi f_{0}$\cite{phif0} resonances. Recently, Particle Data Group categorized it as a vector $s\overline{s}$ meson with a mass of $2.16$ GeV\cite{PDG2020}. In addition, several collaborations have predicted its mass as a $2^{3}D_{1}$ $s\overline{s}$ meson\cite{GI,Ebert,MGI,ERT,QLi}. However, the ground state of D-wave vector $s\overline{s}$ meson $1^{3}D_{1}$ is still not established in experiments. Finally, the predicted masses of $1^{3}D_{1}$ and $2^{3}D_{1}$ $\phi$ mesons by different collaborations were not consistent well with each other\cite{GI,Ebert,MGI,ERT,QLi}.

In order to further confirm the inner structure of $\rho(2000)$ and $\phi(2170)$, we calculate the masses and the decay constants of $1^{3}D_{1}$ and $2^{3}D_{1}$ states of vector D-wave $\rho$ and $\phi$ mesons
based on QCD sum rules. QCD sum rules proved to be a most powerful non-perturbative method in studying the ground state hadrons and
it has been widely used to analyze the masses, decay constants, form factors and strong coupling constants, etc\cite{Shifman,Reinders}.
There have been many reports about its applications to the ground state with spin-parity $J^{PC}=0^{\pm}$, $1^{\pm}$, $2^{+}$, $3^{-}$ mesons\cite{WZG2,Narison,Sundu1,Sundu2,Sundu3,Sundu4,WZG3,YGL2}, while efforts
on the excited states are few\cite{Azizi,WZG4,Van,Hammoud}. In this article, we assign the $\rho(2000)$ and $\phi(2170)$ as the first radially excited  states of D-wave vector $\rho$ and $\phi$ mesons, study
their masses and decay constants with the full QCD sum rules in detail. In our calculations, we consider the contributions of the vacuum condensates
up to dimension-7 in the operator expansion.

The layout of this paper is as follows: in Sec.2, we derive QCD sum rules I(QCDSR I) and II(QCDSR II) from two-point correlation function. QCDSR I and II are respectively used to analyze properties about the ground states of $1^{3}D_{1}$($\rho$), $1^{3}D_{1}$($\phi$) mesons and about the first radially excited states of $2^{3}D_{1}$($\rho$), $2^{3}D_{1}$($\phi$) mesons. Then we present
the numerical results in Sec.3; and Sec.4 is reserved for our conclusions.

\begin{large}
\textbf{2 QCD sum rules for the $\rho(1^{3}D_{1},2^{3}D_{1})$ and $\phi(1^{3}D_{1},2^{3}D_{1})$}
\end{large}

In order to obtain the hadronic observables such as masses, decay constants, we begin our study with the following
two-point correlation function,
\begin{eqnarray}
\notag\
\Pi_{\mu\nu}(p)&&=i\int d^{D}xe^{ip(x-y)}\Big\langle0|\mathbb{T}\Big\{J_{\mu}(x)J_{\nu}^{\dag}(y)\Big\}
|0\Big\rangle\big|_{y\rightarrow0}\\
&&=\Pi(p^{2})(-g_{\mu\nu}+\frac{p_{\mu}p_{\nu}}{p^{2}})+
\Pi_{0}(p^{2})\frac{p_{\mu}p_{\nu}}{p^{2}}
\end{eqnarray}
where $\mathbb{T}$ is the time ordered product and
$J$ is
the interpolating current of vector $\rho$ or $\phi$ meson. The interpolating
current is a composite operator with the same quantum numbers as the
studied hadrons. The current of D-wave vector mesons can be written as,
\begin{eqnarray}
&& J_{\mu}(x)=\overline{q}_{2}(x)\Big(\gamma_{\alpha}\overleftrightarrow{D}_{\mu}\overleftrightarrow{D}_{\alpha}+
\gamma_{\mu}\overleftrightarrow{D}_{\alpha}\overleftrightarrow{D}_{\alpha}+\gamma_{\alpha}\overleftrightarrow{D}_{\alpha}\overleftrightarrow{D}_{\mu} \Big)q_{1}(x)
\end{eqnarray}
\begin{eqnarray}
&& J_{\nu}^{\dag}(y)=\overline{q}_{1}(y)\Big(\gamma_{\beta}\overleftrightarrow{D}_{\mu}\overleftrightarrow{D}_{\beta}+
\gamma_{\nu}\overleftrightarrow{D}_{\beta}\overleftrightarrow{D}_{\beta}+\gamma_{\beta}\overleftrightarrow{D}_{\beta}\overleftrightarrow{D}_{\nu} \Big)q_{2}(y)
\end{eqnarray}
with $\overleftrightarrow{D}_{\mu}=(\overrightarrow{\partial}_{\mu}-ig_{s}G_{\mu})-(\overleftarrow{\partial}_{\mu}+ig_{s}G_{\mu})$, where
$D_{\mu}$, $\partial_{\mu}$ are the covariant derivative, partial derivative and $G_{\mu}$ is the gluon field.
This current can be decomposed
into two parts,
\begin{eqnarray}
J_{\mu}(x)=\eta_{\mu}(x)+J^{V}_{\mu}(x)
\end{eqnarray}
where
\begin{eqnarray}
\eta_{\mu}(x)=&&\overline{q}_{1}(x)\Big(\gamma_{\alpha}\overleftrightarrow{\partial}_{\mu}\overleftrightarrow{\partial}_{\alpha}+
\gamma_{\mu}\overleftrightarrow{\partial}_{\alpha}\overleftrightarrow{\partial}_{\alpha}+\gamma_{\alpha}\overleftrightarrow{\partial}_{\alpha}\overleftrightarrow{\partial}_{\mu} \Big)q_{2}(x) \\
\notag
J^{V}_{\mu}(x)=&&-2i\overline{q}_{2}(x)\Big[\gamma_{\alpha}\Big(g_{s}G_{\mu}\overleftrightarrow{\partial}_{\alpha}+g_{s}G_{\alpha}\overleftrightarrow{\partial}_{\mu}-2ig_{s}^{2}G_{\mu}G_{\alpha}\Big) \\
&& \notag
+\gamma_{\mu}\Big(g_{s}G_{\alpha}\overleftrightarrow{\partial}_{\alpha}+g_{s}G_{\alpha}\overleftrightarrow{\partial}_{\alpha}-2ig_{s}^{2}G_{\alpha}G_{\alpha}\Big) \\
&&
+\gamma_{\alpha}\Big(g_{s}G_{\alpha}\overleftrightarrow{\partial}_{\mu}+g_{s}G_{\mu}\overleftrightarrow{\partial}_{\alpha}-2ig_{s}^{2}G_{\alpha}G_{\mu}\Big) \Big]q_{1}(x)
\end{eqnarray}
and $\overleftrightarrow{\partial}_{\mu}=\overrightarrow{\partial}_{\mu}-\overleftarrow{\partial}_{\mu}$.

This means we have two choices in contructing the interpolating currents of vector D-wave mesons, Eq.(5) with the partial derivative $\partial_{\mu}$\cite{Aliev1,Bagan}, and Eq.(2) with the covariant derivative $D_{\mu}$\cite{Reinders2}. From Ref.\cite{GLY}, we can see that these two currents lead to little differences in the final results in heavy mesons QCD sum rules. To study the effect on light meson, we will present the results coming from the currents with both the partial derivative
and the covariant derivative.

To extract hadronic observables, the correlation will be calculated in two different ways. On one hand, it is identified as as a hadronic propagator, called Phenomenological(Physical) side. On the other hand it is called the QCD side, where it is treated in the framework of the operator product expansion(OPE), and the short and long distance quark-gluon interactions are separated. Then, the result of the QCD calculation is matched, via dispersion relation, to a sum over hadronic states. The sum rule obtained in this way allows to calculate not only observables of the hadronic ground state but also those of excited states.

\begin{large}
\textbf{2.1 The Phenomenological side}
\end{large}

We now describe the first step in the sum rule derivation how the Phenomenological side of the
correlation is done. In this step, a complete set of intermediate hadronic states with the same quantum
numbers as the current operators $J_{\mu}(x)$ is inserted
into the correlation $\Pi_{\mu\nu}(p)$\cite{Shifman,Reinders}. It should be noticed that the interpolating currents
$J_{\mu}(x)$ couples not only to the ground states of hadrons, but also to their excited states with the same quark contents and quantum
numbers,
\begin{eqnarray}
\notag
&&\langle 0| J_{\mu}(0)|\rho/\phi(1D) \rangle=f_{1}M^{2}_{1}\varepsilon_{\mu},  \\
&&\langle 0| J_{\mu}(0)|\rho/\phi(2D) \rangle=f_{2}M^{2}_{2}\widetilde{\varepsilon}_{\mu}
\end{eqnarray}
where $f_{i}$ and $M_{i}$ are the decay constants and the masses of $1^{3}D_{1}$ and $2^{3}D_{1}$ states of $\rho$ or $\phi$ meson, and $\varepsilon_{\mu}$, $\widetilde{\varepsilon}_{\mu}$ are their polarization vectors with the following properties,
\begin{eqnarray}
\widetilde{g}_{\mu\nu}= \mathop{\sum}\limits_\lambda\varepsilon_{\mu}^{*}(\lambda,p)\varepsilon_{\nu}(\lambda,p)
=-g_{\mu\nu}+\frac{p_{\mu}p_{\nu}}{p^{2}}
\end{eqnarray}
In correlation(1), the component $\Pi(p^{2})$ denotes the contributions of the $1^{3}D_{1}$ and $2^{3}D_{1}$ state of vector $\rho$ or $\phi$ meson, while $\Pi_{0}(p^{2})$ comes from the contributions of scalar mesons.  Considering the properties $p^{\mu}\widetilde{g}_{\mu\nu}=p^{\nu}\widetilde{g}_{\mu\nu}=0$, contributions from $1^{3}D_{1}$ and $2^{3}D_{1}$ states can be extracted out by the following projection method,
\begin{eqnarray}
\Pi(p^{2})&&=\frac{1}{3}\widetilde{g}^{\mu\nu}\Pi_{\mu\nu}(p)
\end{eqnarray}
After isolating the pole terms of the
ground state and the first radially excited state, we obtain the following results,
\begin{eqnarray}
\notag\
\Pi_{H}(p^{2})&&=\frac{f_{1}^2M^{4}_{1}}{M_{1}^2-p^2}+
\frac{f_{2}^2M^{4}_{2}}{M_{2}^2-p^2}\cdots \\
&& =\Pi_{\rho/\phi_{(1D)}}(p^2)+\Pi_{\rho/\phi_{(2D)}}(p^2)\cdots
\end{eqnarray}
where $\cdots$ stands for the contributions of the higher resonances
and continuum states, and the subscript $H$ denotes the hadron side of the correlation functions. Then, we obtain the hadronic spectral densities from the imaginary part of correlation function,
\begin{eqnarray}
\notag\
\rho_{H_{1D}}(s)=lim_{\epsilon\rightarrow0}\frac{Im\Pi_{\rho/\phi_{(1D)}}(s+i\epsilon)}{\pi}=f_{1}^2M^{4}_{1}\delta(s-M^{2}_{1}) \\
\rho_{H_{2D}}(s)=lim_{\epsilon\rightarrow0}\frac{Im\Pi_{\rho/\phi_{(2D)}}(s+i\epsilon)}{\pi}=f_{2}^2M^{4}_{2}\delta(s-M^{2}_{2}) \end{eqnarray}

\begin{large}
\textbf{2.2 The QCD side}
\end{large}

We now turn to the next important step in sum rule derivation and describe
how the QCD calculation of the correlation function $(1)$ is done. The correlation function can be approximated at
very large $P^2=-p^2$ by contract all quark fields in Eq.(1) with Wick's theorem. Combined with Eq.(9), the correlation function can be written as,
\begin{eqnarray}
\Pi_{QCD}(p^{2})=&&-\frac{i}{3}\widetilde{g}^{\mu\nu}\int d^{4}xe^{ip.(x-y)}\times\Big\{S^{ji}_{2}(y-x)\Gamma_{\mu}(x)S^{ij}_{1}(x-y)\Gamma_{\nu}(y) \Big\}|_{y=0}
\end{eqnarray}
where $\Gamma_{\mu}(x)$ and $\Gamma_{\nu}(y)$ are the vertexes,
\begin{eqnarray}
\notag\
&& \Gamma_{\mu}(x)=\gamma_{\alpha}\overleftrightarrow{D}_{\mu}\overleftrightarrow{D}_{\alpha}+
\gamma_{\mu}\overleftrightarrow{D}_{\alpha}\overleftrightarrow{D}_{\alpha}+\gamma_{\alpha}\overleftrightarrow{D}_{\alpha}\overleftrightarrow{D}_{\mu} \end{eqnarray}
\begin{eqnarray}
\notag\
&& \Gamma_{\nu}(y)=\gamma_{\beta}\overleftrightarrow{D}_{\mu}\overleftrightarrow{D}_{\beta}+
\gamma_{\nu}\overleftrightarrow{D}_{\beta}\overleftrightarrow{D}_{\beta}+\gamma_{\beta}\overleftrightarrow{D}_{\beta}\overleftrightarrow{D}_{\nu} \end{eqnarray}
and $S$ are the quark propagators in the coordinate or momentum spaces which can be written as,
\begin{eqnarray}
 S_{q}^{ij}(x)=&&i\frac{x\!\!\!/}{2\pi^{2}x^{4}}\delta_{ij}-\frac{m_{q}}{4\pi^{2}x^{2}}\delta_{ij}-\frac{\langle
 \overline{q}q\rangle}{12}\Big(1-i\frac{m_{q}}{4}x\!\!\!/\Big)-\frac{x^{2}}{192}m_{0}^{2}\langle
 \overline{q}q\rangle\Big( 1-i\frac{m_{q}}{6}x\!\!\!/\Big)\\
 &&
 \notag-\Big[\frac{\lambda^{a}}{2}\Big]^{ij}\Big[x\!\!\!/\sigma^{\theta\eta}+\sigma^{\theta\eta}x\!\!\!/\Big]\frac{i}{32\pi^{2}x^{2}}g_{s}G^{a}_{\theta\eta}+\frac{i\delta_{ij}x^{2}x\!\!\!/g_{s}^{2}\langle \overline{q}q\rangle^{2}}{7776}-\frac{\delta_{ij}x^{4}\langle \overline{q}q\rangle\langle GG\rangle}{27648}\cdots,
\end{eqnarray}
\begin{eqnarray}
\notag
 S_{Q}^{ij}(x)=&&\frac{i}{(2\pi)^{4}}\int
 d^{4}ke^{-ik.x}\Big\{\frac{\delta_{ij}}{k\!\!\!/-m_{Q}}-\frac{g_{s}G_{\alpha\beta}^{c}t^{c}_{ij}}{4}\frac{\sigma^{\alpha\beta}(k\!\!\!/+m_{Q})+(k\!\!\!/+m_{Q})\sigma^{\alpha\beta}}{(k^{2}-m_{Q}^{2})^{2}}\\
 &&-\frac{g_{s}^{2}(t^{a}t^{b})_{ij}G_{\alpha\beta}^{a}G_{\mu\nu}^{b}(f^{\alpha\beta\mu\nu}+f^{\alpha\mu\beta\nu}+f^{\alpha\mu\nu\beta})}{4(k^{2}-m_{Q}^{2})^{5}}\\
 \notag &&
 + \frac{i\langle g_{s}^{3}GGG\rangle}{48}\frac{(k\!\!\!/+m_{Q})\big[k\!\!\!/(k^{2}-3m_{Q}^{2})+2m_{Q}(2k^{2}-m_{Q}^{2})\big](k\!\!\!/+m_{Q})}{(k^{2}-m_{Q}^{2})^{6}}\cdots\Big\}
\end{eqnarray}
where
\begin{eqnarray}
f^{\alpha\beta\mu\nu}=(k\!\!\!/+m_{Q})\gamma^{\alpha}(k\!\!\!/+m_{Q})\gamma^{\beta}(k\!\!\!/+m_{Q})\gamma^{\mu}(k\!\!\!/+m_{Q})\gamma^{\nu}(k\!\!\!/+m_{Q})
\end{eqnarray}
In the fixed point gauge, $G_{\mu}(x)=\frac{1}{2}x^{\theta}G_{\theta\mu}(0)+\cdots$ and $G_{\alpha}(y)=\frac{1}{2}y^{\theta}G_{\theta\alpha}(0)+\cdots|_{y=0}=0$.
Thus, for the vertex $\Gamma_{\nu}(y)$, we can get $G_{\alpha}(y)=G_{\mu}(y)=0$, there are no gluon lines associated with the vertex at the point $y=0$.
After completing the integrals both in the coordinate and momentum spaces, we obtain the QCD spectral density through the imaginary part of the correlation,
\begin{eqnarray}
\rho_{QCD}(s)=lim_{\epsilon\rightarrow0}\frac{Im\Pi_{QCD}(s+i\epsilon)}{\pi}
\end{eqnarray}
After lengthy derivation, we find there is no contribution of the condensate terms $\langle \overline{q}q\rangle$, $\langle \overline{q}g_{s}\sigma Gq\rangle$ and $\langle \overline{q}q\rangle^{2}$. For $\phi$ meson, its spectral densities of perturbative and non-perturbative terms are listed in Eqs.(17)$-$(19), where $\rho^{V}_{\phi-nonpert}(s)$ denotes the contributions from vertex and $\rho_{\phi-nonpert}^{\eta}(s)$ represents the other contributions in the non-perturbative terms. For $\rho$ meson, because its spectral densities of non-perturbative terms are tediously lengthy, we only list its perturbative part in Eq.(20),
\begin{eqnarray}
\rho_{\phi-pert}(s)=\frac{1}{4\pi^{2}}\times\frac{(s-4m^{2})^{\frac{5}{2}}(s+18m^{2})}{\sqrt{s}}
\end{eqnarray}
\begin{eqnarray}
\notag\
\rho_{\phi-nonpert}^{\eta}(s)&&=\Big[\frac{\sqrt{s(s-4m^{2})}(162m^{4}+421m^{2}s-136s^{2})}{36s^{2}}+\frac{m^{2}(320m^{4}-944m^{2}s+241s^{2})}{10s^2} \Big]\\
&&\times\langle\frac{\alpha_{s}GG}{\pi}\rangle+\Big[\frac{-5184m^{8}+15972m^{6}s+7804m^{4}s^{2}-5574m^{2}s^{3}+693s^{4}}{1152\pi^{2}s^{2}(4m^{2}-s)\sqrt{s(s-4m^{2})}} \\
\notag\
&&+\frac{368m^{4}-384m^{2}s+49s^{2}}{384\pi^{2}s^{2}}\Big] \times\langle g_{s}^{3}GGG\rangle -\Big[\frac{17}{9}+\frac{87m^{2}}{108T^2}+\frac{7m^{4}}{18T^4}+\frac{m^{6}}{24T^6}+\frac{m^{8}}{144T^8}\Big]\\
\notag\
&&\times m<qqGG>
\end{eqnarray}
\begin{eqnarray}
\notag
\rho^{V}_{\phi-nonpert}(s)&&= \Big[-\frac{1}{6}m^{2}+\frac{s(3s-8m^{2})}{4\sqrt{s(s-4m^{2})}}-\frac{m^{4}s(5s-8m^{2})}{9[s(s-4m^{2})]^{\frac{3}{2}}} +\frac{3s(s-2m^{2})}{4\sqrt{s(s-4m^{2})}}\\
\notag\
&&-\frac{2m^{4}s(s-m^{2})}{3[s(s-4m^{2})]^{\frac{3}{2}}}\Big]\times\langle\frac{\alpha_{s}GG}{\pi}\rangle
+\frac{1}{3}\Big[\frac{s+12m^{2}}{4\sqrt{s(s-4m^{2})}}+\frac{m^{2}-6s}{4s}\\
&&
+\frac{m^{4}-2m^{2}s-s^{2}}{8(s-m^{2})^{2}}
\Big]\times\langle g_{s}^{3}GGG\rangle
\end{eqnarray}
\begin{eqnarray}
\notag\
\rho_{\rho-pert}(s)&&=\frac{\sqrt{[s-(m_{1}-m_{2})^{2}][s-(m_{1}+m_{2})^{2}]}}{8\pi^{2}s} \\
\notag\
&&\times\Big[16_{1}^{8}+32m_{1}^{7}m_{2}+16m_{1}^{6}(m^{2}_{2}-2s)-8m_{1}^{5}(4m_{2}^{3}+13m_{2}s)\\
&&+m_{1}^{4}(-64m_{2}^{4}-80m^{2}_{2}s+17s^{2})-m^{3}_{1}(4m_{2}^{5}+18m_{2}^{3}s-11m_{2}s^{2}) \\
\notag\
&&+m_{1}^{2}(16m_{2}^{6}-80m_{2}^{4}s+46m_{2}^{2}s^{2}+s^{3})
+2m_{1}(16m_{2}^{7}-52m_{2}^{5}s+44m_{2}^{3}s^{2}-11m_{2}s^{3})\\
\notag\
&&
+16m_{2}^{8}-32m^{6}_{2}s+17m_{2}^{4}s^{2}+m^{2}_{2}s^{3}-2s^{4}\Big]
\end{eqnarray}
Using dispersion relations, the correlation function can be written as the following form,
\begin{eqnarray}
\Pi_{QCD}(p^{2})=\int^{\infty}_{(m_{1}+m_{2})^{2}}\frac{\rho_{QCD}(s)}{s-p^{2}}ds
\end{eqnarray}
We take quark-hadron duality and perform the Borel transformation to the Phenomenological side as well as the QCD side
to obtain the QCD sum rules,
\begin{eqnarray}
\notag\
&&f_{1}^2M^{4}_{1}exp\Big[-\frac{M^{2}_{1}}{T^2}\Big]\\
\notag\
=&&\int_{(m_{1}+m_{2})^{2}}^{s_{0}}\big[\rho_{H_{1D}}(s)+\rho_{H_{2D}}(s)\big]exp\Big[-\frac{s}{T^2}\Big]ds\\
=&&\int_{(m_{1}+m_{2})^{2}}^{s_{0}}\rho_{QCD}(s)exp\Big[-\frac{s}{T^2}\Big]ds
\end{eqnarray}
\begin{eqnarray}
\notag\
&&f_{1}^2M^{4}_{1}exp\Big[-\frac{M^{2}_{1}}{T^2}\Big]+f_{2}^2M^{4}_{2}exp\Big[-\frac{M^{2}_{2}}{T^2}\Big]
\\
\notag\
&&=\int_{(m_{1}+m_{2})^{2}}^{s'_{0}}\big[\rho_{H_{1D}}(s)+\rho_{H_{2D}}(s)\big]exp\Big[-\frac{s}{T^2}\Big]ds\\
&&=\int_{(m_{1}+m_{2})^{2}}^{s'_{0}}\rho_{QCD}(s)exp\Big[-\frac{s}{T^2}\Big]ds
\end{eqnarray}
In Eq.(22), $s_{0}$ is the continuum threshold parameter, which separates the contribution of the ground state $1^{3}D_{1}$ from those of the higher resonances and continuum states. While $s'_{0}$ separates the contribution of $1^{3}D_{1}$ plus $2^{3}D_{1}$ states from the higher resonances and continuum states. By differentiating Eq.(22) with respect to $\frac{1}{T^{2}}$, and eliminating the decay constant $f_{1}^{2}$, we obtain,
\begin{eqnarray}
M_{1}^{2}=\frac{-\frac{d}{d(1/T^{2})}\int_{(m_{1}+m_{2})^{2}}^{s_{0}}\rho_{QCD}(s)exp\Big[-\frac{s}{T^2}\Big]ds}{\int_{(m_{1}+m_{2})^{2}}^{s_{0}}\rho_{QCD}(s)exp\Big[-\frac{s}{T^2}\Big]ds}
\end{eqnarray}
After the mass $M_{1}^{2}$ is obtained, it is treated as a input parameter to obtain the decay constant from QCD sum rule from Eq.(22),
\begin{eqnarray}
f_{1}^{2}=\frac{\int_{(m_{1}+m_{2})^{2}}^{s_{0}}\rho_{QCD}(s)exp\Big[-\frac{s}{T^2}\Big]ds}{M^{4}_{1}exp\Big[-\frac{M_{1}^2}{T^2}\Big]}
\end{eqnarray}
In the following, we will refer to the QCD sum rules in Eq.(24) and Eq.(25) as QCDSR I.

We introduce the notations $\tau=\frac{1}{T^{2}}$, $D^{n}=(-\frac{d}{d\tau})^{n}$ and $\Pi'_{QCD}(\tau)=\int_{(m_{1}+m_{2})^{2}}^{s'_{0}}\rho_{QCD}(s)exp\Big[-\tau s\Big]ds$ for simplicity.  With these substitutions, Eq(23) can be written as,
\begin{eqnarray}
f_{1}^2M^{4}_{1}exp\Big[-\tau M^{2}_{1}\Big]+f_{2}^2M^{4}_{2}exp\Big[-\tau M^{2}_{2}\Big]=\Pi'_{QCD}(\tau)
\end{eqnarray}
Then, deriving the QCD sum rules in Eq.(26) with respect to $\tau$, we obtain,
\begin{eqnarray}
f_{1}^2M^{6}_{1}exp\Big[-\tau M^{2}_{1}\Big]+f_{2}^2M^{6}_{2}exp\Big[-\tau M^{2}_{2}\Big]=D\Pi'_{QCD}(\tau)
\end{eqnarray}
From Eq.(26) and Eq.(27), we can obtain the following relation,
\begin{eqnarray}
f_{i}^{2}exp(-\tau M_{i}^{2})=\frac{(D-M_{j}^{2})\Pi'_{QCD}(\tau)}{M_{i}^{4}(M_{i}^{2}-M_{j}^{2})}
\end{eqnarray}
After deriving with respect to $\tau$ in Eq.(28), we obtain the following two relations,
\begin{eqnarray}
M_{i}^{2}=\frac{(D^{2}-M_{j}^{2}D)\Pi'_{QCD}(\tau)}{(D-M_{j}^{2})\Pi'_{QCD}(\tau)}
\end{eqnarray}
\begin{eqnarray}
M_{i}^{4}=\frac{(D^{3}-M_{j}^{2}D^{2})\Pi'_{QCD}(\tau)}{(D-M_{j}^{2})\Pi'_{QCD}(\tau)}
\end{eqnarray}
From these two relations, we can get a equation about the squared masses $M_{i}^{2}$,
\begin{eqnarray}
M_{i}^{4}-bM_{i}^{2}+c=0
\end{eqnarray}
where
\begin{eqnarray}
b=\frac{D^{3}\otimes D^{0}-D^{2}\otimes D}{D^{2}\otimes D^{0}-D\otimes D},
\end{eqnarray}
\begin{eqnarray}
c=\frac{D^{3}\otimes D-D^{2}\otimes D^{2}}{D^{2}\otimes D^{0}-D\otimes D},
\end{eqnarray}
\begin{eqnarray}
D^{j}\otimes D^{k}=D^{j}\Pi'_{QCD}(\tau)D^{k}\Pi'_{QCD}(\tau),
\end{eqnarray}
the indexes $i=1,2$ and $j,k=0,1,2,3$. By solving the equation in Eq.(31) analytically, we finally obtain two solutions\cite{Maior,ZGWang5,ZGWang6},
\begin{eqnarray}
M_{1}^{2}=\frac{b-\sqrt{b^{2}-4c}}{2}
\end{eqnarray}
\begin{eqnarray}
M_{2}^{2}=\frac{b+\sqrt{b^{2}-4c}}{2}
\end{eqnarray}
In Eqs.(35)-(36), $M_{1}$ and $M_{2}$ represent the masses of the ground state and the first radially excited state. Because the ground state mass in Eq.(35) suffers from additional uncertainties from the first radially excited state $2^{3}D_{1}$, we neglect this result and still use QCDSR I to obtain the mass and decay constant of ground state. In the following, we will refer to the QCD sum rules in Eq.(28) and Eq.(36) as the QCDSR II which is used to analyze the properties of the first radial excitation.

\begin{large}
\textbf{2.3 The numerical results}
\end{large}

In the QCD side, we take the the vacuum condensates to be the standard values $\langle \overline{q}q\rangle=-(0.24\pm0.01$ GeV$)^{3}$, $\langle \overline{s}s\rangle=(0.8\pm0.1)\langle \overline{q}q\rangle$, $\langle\frac{\alpha_{s}GG}{\pi}\rangle=(0.012\pm0.004)$ GeV$^{4}$, $\langle g_{s}^{3}GGG\rangle=0.045$ GeV$^{6}$\cite{Shifman,Reinders,Colangelo}. And the masses of quarks are taken to be $m_{u}=2.16^{+0.49}_{-0.26}$ MeV, $m_{d}=4.67^{+0.48}_{-0.17}$ MeV and $m_{s}=93^{+11}_{-5}$ MeV from the  Particle Data Group\cite{PDG2020}. The final results also depend on two parameters, the Borel mass parameter $T^{2}$ and continuum threshold $s_{0}$($s'_{0}$). In order to choose the working interval of the parameters $T^{2}$ and
$s_{0}$($s'_{0}$), two criteria should be satisfied which are pole dominance and OPE convergence. That is to say, the pole contribution should be as large as possible(larger than $40\%$) comparing with the contributions of the high resonances and continuum states. Meanwhile, we
should also find a plateau, which will ensure OPE convergence and the stability
of the final results. The plateau is often called Borel window.
\begin{figure}[h]
\begin{minipage}[t]{0.45\linewidth}
\centering
\includegraphics[height=5cm,width=7cm]{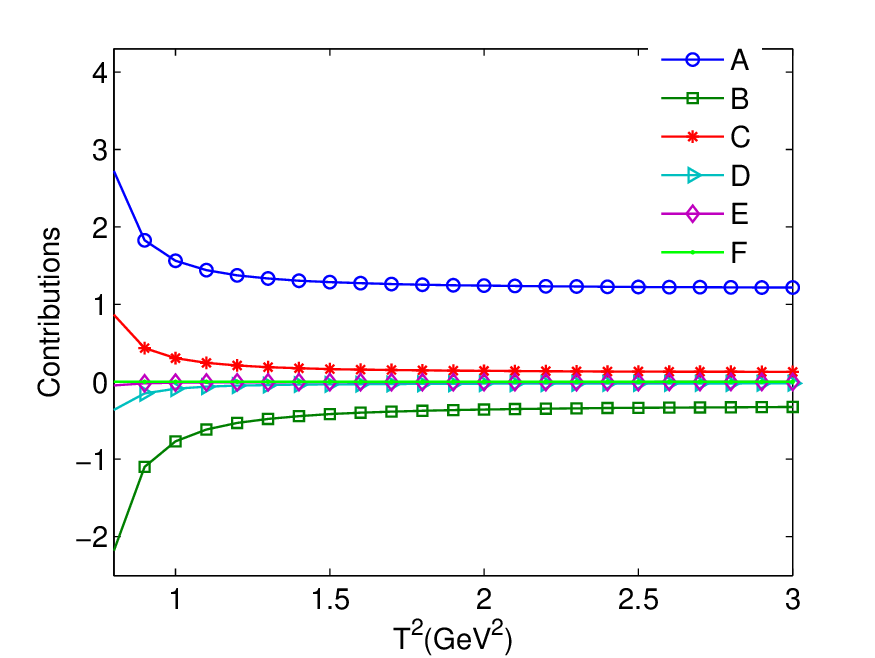}
\caption{The contributions of different condensate terms in the OPE with variations of the Borel
parameters $T^2$ for $\rho$($1^{3}D_{1}$) meson, where A-F denote perturbative term, $\langle \frac{\alpha_{s}GG}{\pi}\rangle_{\eta_{\mu}}$, $\langle \frac{\alpha_{s}GG}{\pi}\rangle_{J^{V}_{\mu}}$, $\langle g_{s}^{3}GGG\rangle_{\eta_{\mu}}$, $\langle g_{s}^{3}GGG\rangle_{J^{V}_{\mu}}$, and $\langle q\overline{q}GG\rangle$.\label{your
label}}
\end{minipage}
\hfill
\begin{minipage}[t]{0.45\linewidth}
\centering
\includegraphics[height=5cm,width=7cm]{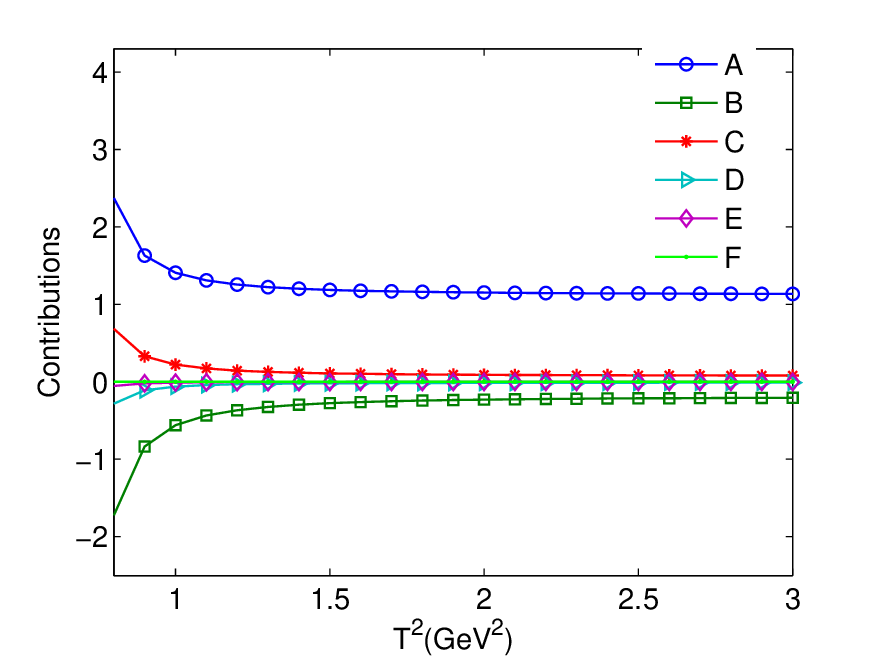}
\caption{The contributions of different condensate terms in the OPE with variations of the Borel
parameters $T^2$ for $\rho$($2^{3}D_{1}$) meson, where A-F denote perturbative term, $\langle \frac{\alpha_{s}GG}{\pi}\rangle_{\eta_{\mu}}$, $\langle \frac{\alpha_{s}GG}{\pi}\rangle_{J^{V}_{\mu}}$, $\langle g_{s}^{3}GGG\rangle_{\eta_{\mu}}$, $\langle g_{s}^{3}GGG\rangle_{J^{V}_{\mu}}$, and $\langle q\overline{q}GG\rangle$.\label{your
label}}
\end{minipage}
\end{figure}

\begin{figure}[h]
\begin{minipage}[t]{0.45\linewidth}
\centering
\includegraphics[height=5cm,width=7cm]{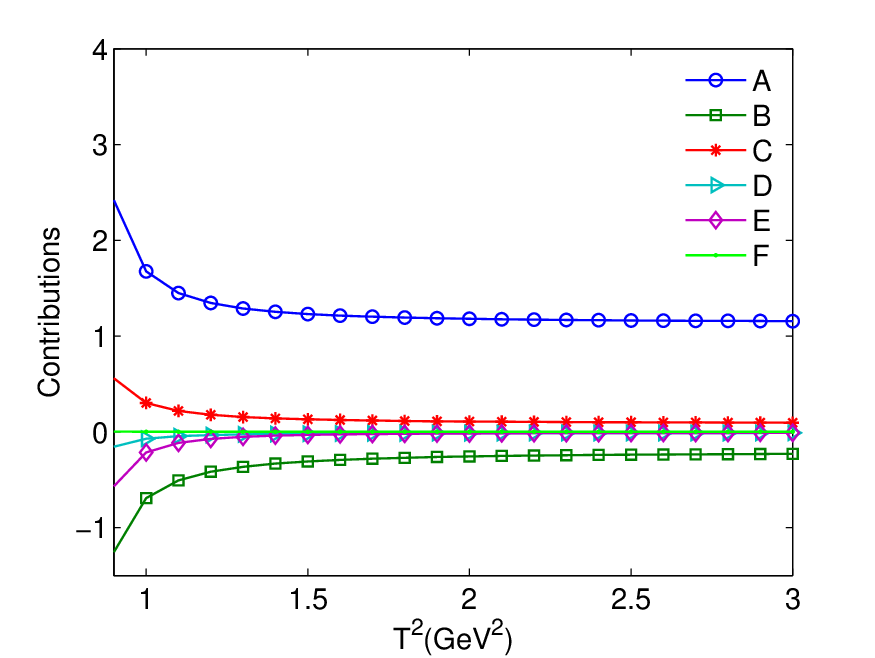}
\caption{The contributions of different condensate terms in the OPE with variations of the Borel
parameters $T^2$ for $\phi$($1^{3}D_{1}$) meson, where A-F denote perturbative term, $\langle \frac{\alpha_{s}GG}{\pi}\rangle_{\eta_{\mu}}$, $\langle \frac{\alpha_{s}GG}{\pi}\rangle_{J^{V}_{\mu}}$, $\langle g_{s}^{3}GGG\rangle_{\eta_{\mu}}$, $\langle g_{s}^{3}GGG\rangle_{J^{V}_{\mu}}$, and $\langle q\overline{q}GG\rangle$.\label{your
label}}
\end{minipage}
\hfill
\begin{minipage}[t]{0.45\linewidth}
\centering
\includegraphics[height=5cm,width=7cm]{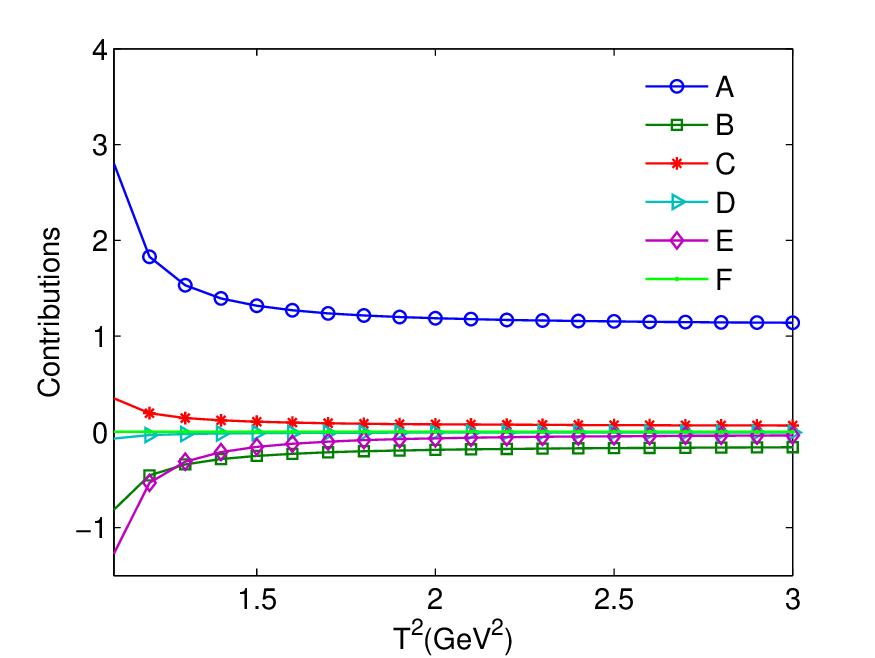}
\caption{The contributions of different condensate terms in the OPE with variations of the Borel
parameters $T^2$ for $\phi$($2^{3}D_{1}$) meson, where A-F denote perturbative term, $\langle \frac{\alpha_{s}GG}{\pi}\rangle_{\eta_{\mu}}$, $\langle \frac{\alpha_{s}GG}{\pi}\rangle_{J^{V}_{\mu}}$, $\langle g_{s}^{3}GGG\rangle_{\eta_{\mu}}$, $\langle g_{s}^{3}GGG\rangle_{J^{V}_{\mu}}$, and $\langle q\overline{q}GG\rangle$.\label{your
label}}
\end{minipage}
\end{figure}
The contributions from different parts of spectra density are defined as $\frac{\rho_{D_{i}}}{\sum \rho _{D_{i}}}$, where $D_{i}$ represents the dimension of condensate terms. In Figs.1-4, we show the dependence of these condensate terms on Borel parameter $T^{2}$. From these figures, we can see good OPE convergence for both $\rho$ and $\phi$ mesons. When the Borel parameters are larger than $1.1$ GeV$^{2}$ for $\rho$($1^{3}D_{1}$), $\rho$($2^{3}D_{1}$) and $\phi$($1^{3}D_{1}$) and larger than $1.4$ GeV$^{2}$ for $\phi$($2^{3}D_{1}$), the results are rather stable with variations of the Borel parameters and the contributions from perturbative term is larger than $60\%$(see Table I).

The threshold parameters $s_{0}$ and $s'_{0}$ are used to include the ground state and the ground state plus the first radially excited state, respectively. Commonly, these parameters are chosen to be $\sqrt{s_{0}}=M_{1}+\Delta_{1}$ and $\sqrt{s'_{0}}=M_{2}+\Delta_{2}$. In order to include contributions from the ground state and to exclude the contaminations from the first radial excitation, the value of $\Delta_{1}$ should be little than the gap between $1^{3}D_{1}$ state and $2^{3}D_{1}$ state. For $\rho$ meson as an example, the masses of $1^{3}D_{1}$ state and $2^{3}D_{1}$ state are predicted to be about $1.7$ GeV and $2.0$ GeV\cite{Bugg}, respectively. Theoretically, the value of $\Delta_{1}$ is $\sim 0.3$ GeV. However, we should also consider its total width of $2^{3}D_{1}$ state which was predicted to be ($234\pm30\pm25$) MeV in reference\cite{BESIII}. Thus, the value of $\Delta_{1}$ for $\rho$ meson should be little than $0.3$ GeV. In Fig.5, we show dependence of the mass of $1^{3}D_{1}$ $\rho$ meson on Borel parameters $T^{2}$ in different values of $\Delta_{1}$, where $\sqrt{s_{0}}=M_{1}+\Delta_{1}=1.7+0.1=1.8$ GeV, $1.7+0.2=1.9$ GeV, $1.7+0.3=2.0$ GeV. When the Borel parameters are larger than $1.0$ GeV$^{2}$, we can see from Fig.5 that the curve with $\sqrt{s_{0}}=M_{1}+\Delta_{1}=1.7+0.2=1.9$ GeV shows more stability than those curves with $\sqrt{s_{0}}=1.8$ GeV and $\sqrt{s_{0}}=2.0$ GeV. In addition, the curve change rapidly with variation of the Borel parameters if $\sqrt{s_{0}}$ is taken to be $1.7+0.3=2.0$ GeV. This is because of the contaminations of the first radially excited state $2^{3}D_{1}$. Combined with these above considerations, the working region for the Borel parameter and continuum threshold for $\rho$($1^{3}D_{1}$) meson are determined  to be $1.1-1.5$ GeV$^2$ and $\sqrt{s_{0}}=1.9$ GeV which are presented in Table I together with the pole and purturbative contributions.

\begin{figure}[h]
\begin{minipage}[t]{0.45\linewidth}
\centering
\includegraphics[height=5cm,width=7cm]{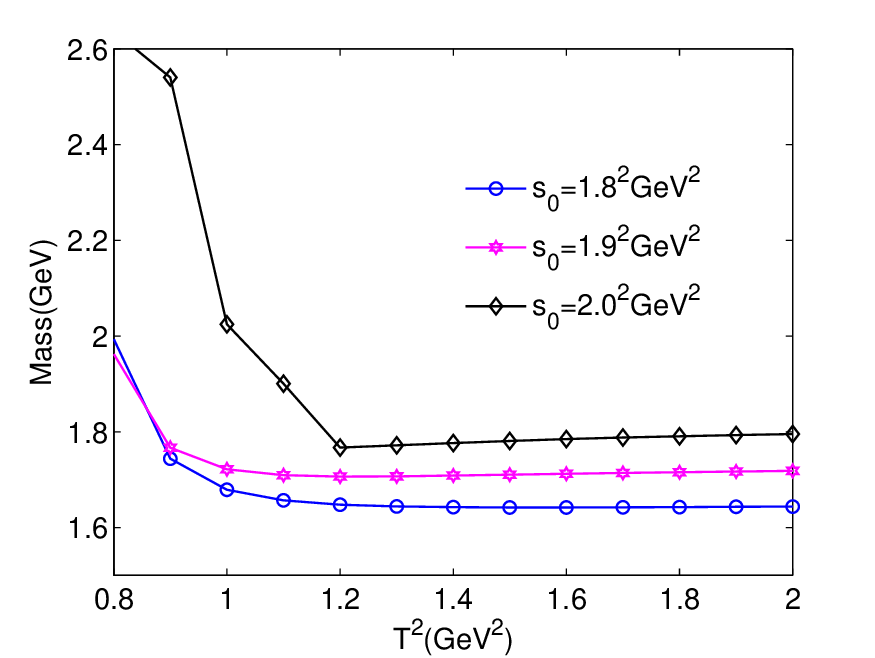}
\caption{The results for $\rho$($1^{3}D_{1}$) meson on Borel
parameter $T^{2}$, with different values of $s_{0}$ in QCDSR I.\label{your label}}
\end{minipage}
\hfill
\begin{minipage}[t]{0.45\linewidth}
\centering
\includegraphics[height=5cm,width=7cm]{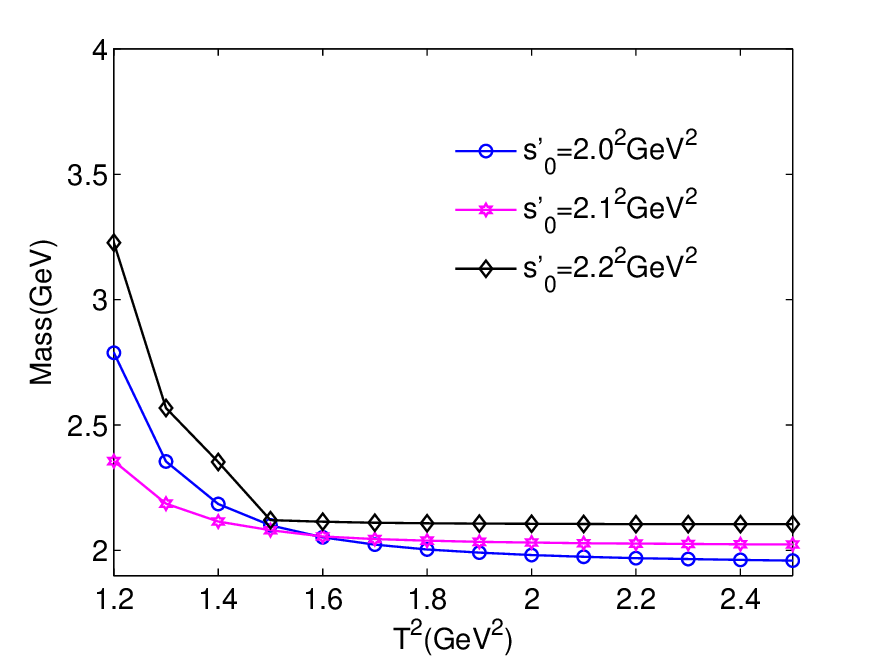}
\caption{The results for $\rho$($2^{3}D_{1}$) meson on Borel
parameter $T^{2}$, with different values of $s'_{0}$ in QCDSR II.\label{your label}}
\end{minipage}
\end{figure}
In reference\cite{LPHe,Bugg}, the mass and decay width of the second radial excitation $\rho$($3^{3}D_{1}$) were predicted to be $\sim2.27$ GeV and $\sim300$ MeV. The energy gap between the first and the second radially excited states is about $250$ MeV which suggests the continuum threshold parameter for $\rho$($2^{3}D_{1}$) in QCDSR II is $\sqrt{s'_{0}}\approx2.0+0.25=2.25$ GeV. Considering the width of the second radially excited state, the value of $\sqrt{s'_{0}}$ in QCDSR II should be little than $2.25$ GeV. From Fig.6, we can see that the results are more stable with continuum parameter $\sqrt{s'_{0}}=2.1$ GeV than those of $\sqrt{s'_{0}}=2.0$ GeV and $\sqrt{s'_{0}}=2.2$ GeV. This phenomenon indicates that too large values of $s'_{0}$ will lead to contaminations from the second radially excited state and too small values can not totally include contribution of the first radially excitation. This above conclusion about $\rho$ meson is also applicable to $\phi$ meson. In reference\cite{GI,Ebert,MGI,ERT}, we know that the masses of the ground state, the first and the second radially excited states
of vector $\phi$ meson are $\sim1.87$ GeV, $\sim2.3$ GeV and $\sim2.6$ GeV. Their strong decay widths are $\sim400$ MeV, $\sim200$ MeV and $\sim200$ MeV respectively. Combining with results obtained in different values of $s_{0}$($s'_{0}$) which are shown in Figs.7-8, we determine the working regions of Borel parameter $T^{2}$ and thresholds $s_{0}$($s'_{0}$) and also present them in Table I.
\begin{figure}[h]
\begin{minipage}[t]{0.45\linewidth}
\centering
\includegraphics[height=5cm,width=7cm]{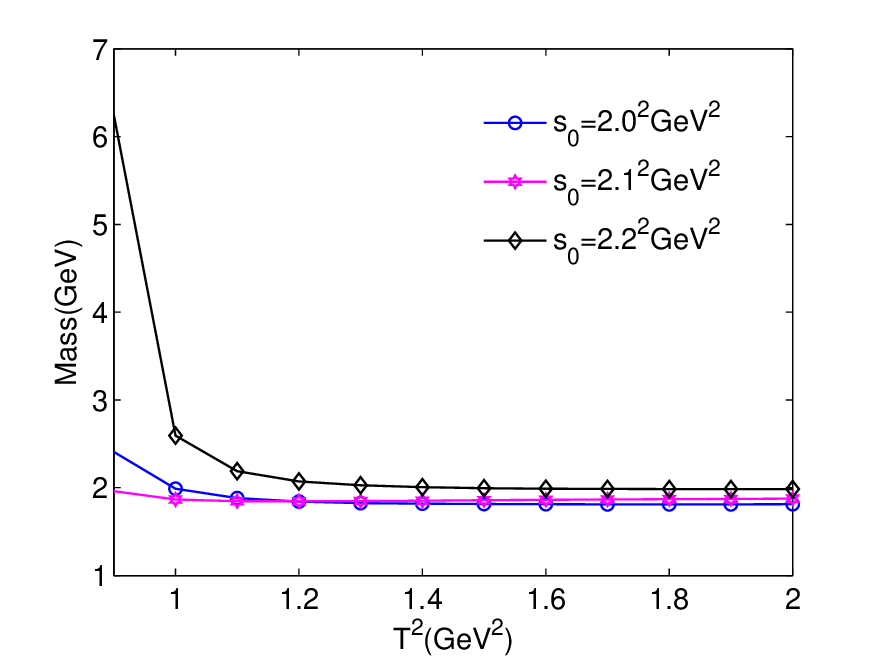}
\caption{The results for $\phi$($1^{3}D_{1}$) meson on Borel
parameter $T^{2}$, with different values of $s_{0}$ in QCDSR I.\label{your label}}
\end{minipage}
\hfill
\begin{minipage}[t]{0.45\linewidth}
\centering
\includegraphics[height=5cm,width=7cm]{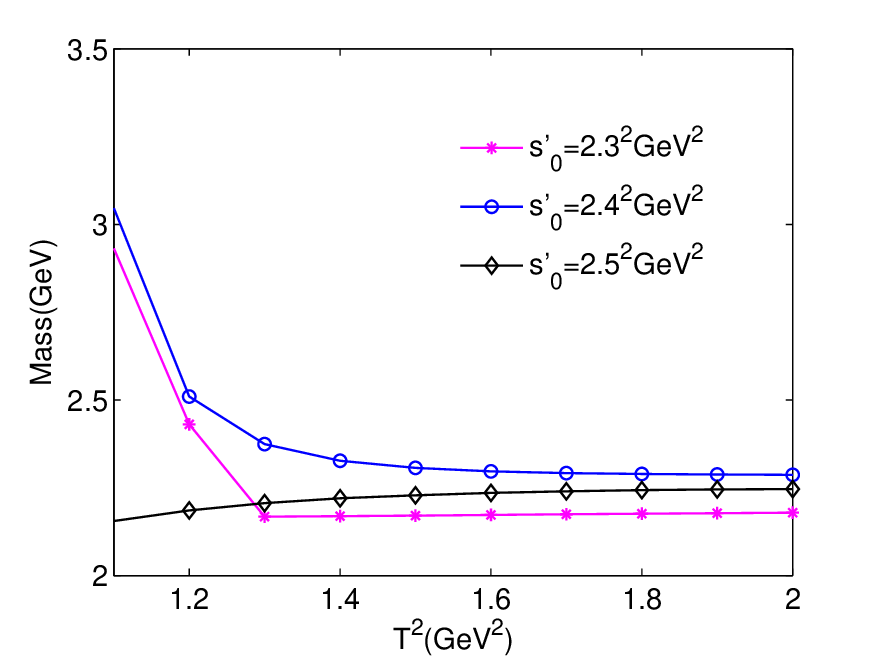}
\caption{The results for $\phi$($2^{3}D_{1}$) meson on Borel
parameter $T^{2}$, with different values of $s'_{0}$ in QCDSR II.\label{your label}}
\end{minipage}
\end{figure}

\begin{table*}[htbp]
\begin{ruledtabular}\caption{The Borel parameters $T$$^{2}$
and continuum threshold parameters $s_{0}$($s'_{0}$) for $\rho$ and $\phi$ meson, where
the pole stands for the pole contributions from the ground states or the ground states plus the first radially
excited states, and the perturbative stands for the absolute value of contributions from the perturbative
terms.}
\begin{tabular}{c| c c c c c c}
  State& \ $T^{2}$(GeV$^{2}$)  & \ $\sqrt{s_{0}}$(GeV) &\ pole  &\ perturbative  \\
 \hline
$\rho(1^{3}D_{1})$ &\  $1.1-1.5$      &\  $1.9$   &\  $40-44\%$   &\ $61-68\%$  \\
$\rho(2^{3}D_{1})$ &\  $1.5-1.9$      &\  $2.1$   &\  $42-50\%$  & \  $67-74\%$   \\
$\phi(1^{3}D_{1})$&\  $1.1-1.5$     &\  $2.1$   &\ $43-46\%$  & \  $62-71\%$   \\
$\phi(2^{3}D_{1})$ &\  $1.4-1.8$     &\  $2.3$   &\ $45-50\%$ &\  $69-76\%$    \\
\end{tabular}
\end{ruledtabular}
\end{table*}

After two criteria of QCD sum rules are both satisfied, we can easily extract physical quantities. Before these extraction, we give a simple discussion about effects of different currents defined by Eqs.(2) and (5) on the final results. It can be seen from Figs.9-12 that the contribution of vertex $J^{V}_{\mu}$ in $J_{\mu}$ has a significant influence on the masses. These figures clearly show that the values coming from current $J_{\mu}=\eta_{\mu}+J^{V}_{\mu}$ are more stable in the Borel window than those coming from $\eta_{\mu}$. That is,  we can not obtain ideal Borel window if contribution coming from vertex $J_{\mu}^{V}$ was not considered. This effect can easily be explained according to Figs.1-4, which show that the contributions of vertex $J^{V}_{\mu}$ mainly originate from the condensate term
$\langle\frac{\alpha_{s}}{\pi}GG\rangle_{J^{V}_{\mu}}$. This condensate term from vertex is at the same order of magnitude with $\langle\frac{\alpha_{s}}{\pi}GG\rangle$ and its contribution makes up $5\%-8\%$ of the total contributions. Thus, its influence on the final results is obvious. In addition, contribution of $\langle g_{s}^{3}GGG\rangle_{J^{V}_{\mu}}$ is one order of magnitude lower than that of $\langle\frac{\alpha_{s}}{\pi}GG\rangle_{J^{V}_{\mu}}$, but its contribution is comparable with that of $\langle g_{s}^{3}GGG\rangle$. Thus, all these condensate terms coming from vertex play an important role and should not be neglected in light meson QCD sum rules.
\begin{figure}[h]
\begin{minipage}[t]{0.45\linewidth}
\centering
\includegraphics[height=5cm,width=7cm]{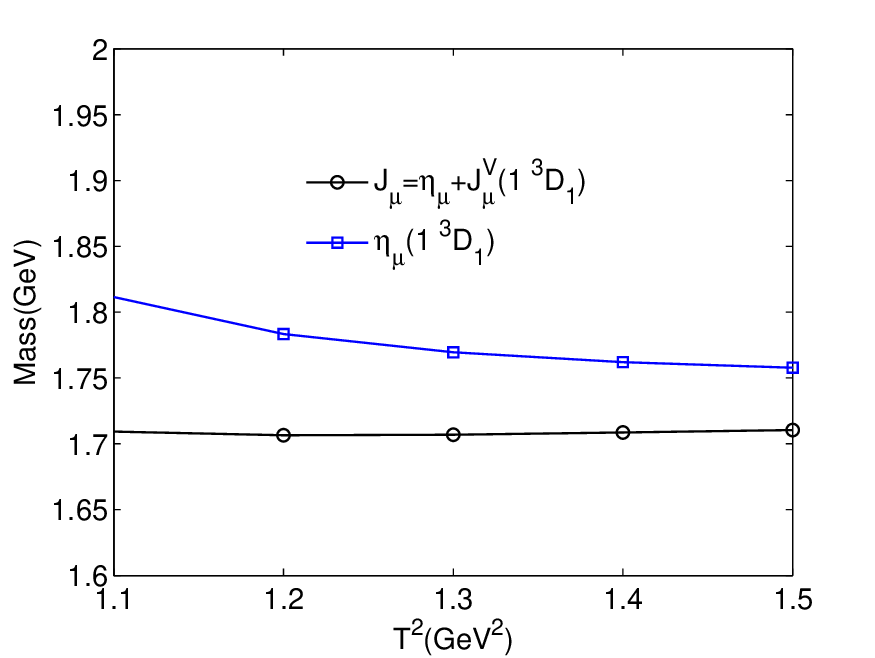}
\caption{The mass of $\rho$($1^{3}D_{1}$) state for different values
of the Borel parameter $T^{2}$
, where the currents
$J_{\mu}$ and $\eta_{\mu}$ are considered respectively.\label{your label}}
\end{minipage}
\hfill
\begin{minipage}[t]{0.45\linewidth}
\centering
\includegraphics[height=5cm,width=7cm]{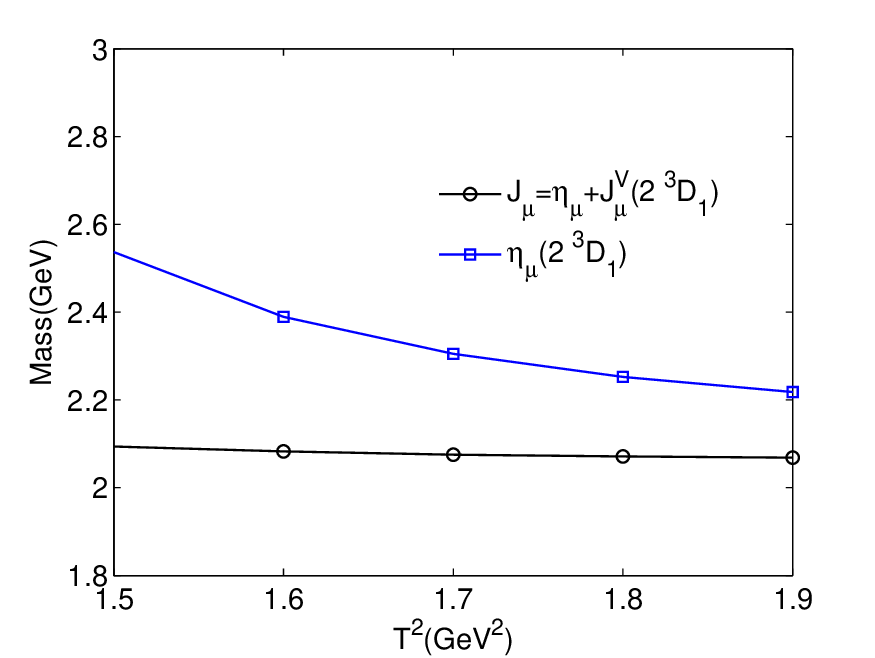}
\caption{The mass of $\rho$($2^{3}D_{1}$) state for different values
of the Borel parameter $T^{2}$
, where the currents
$J_{\mu}$ and $\eta_{\mu}$ are considered respectively.\label{your label}}
\end{minipage}
\end{figure}
\begin{figure}[h]
\begin{minipage}[t]{0.45\linewidth}
\centering
\includegraphics[height=5cm,width=7cm]{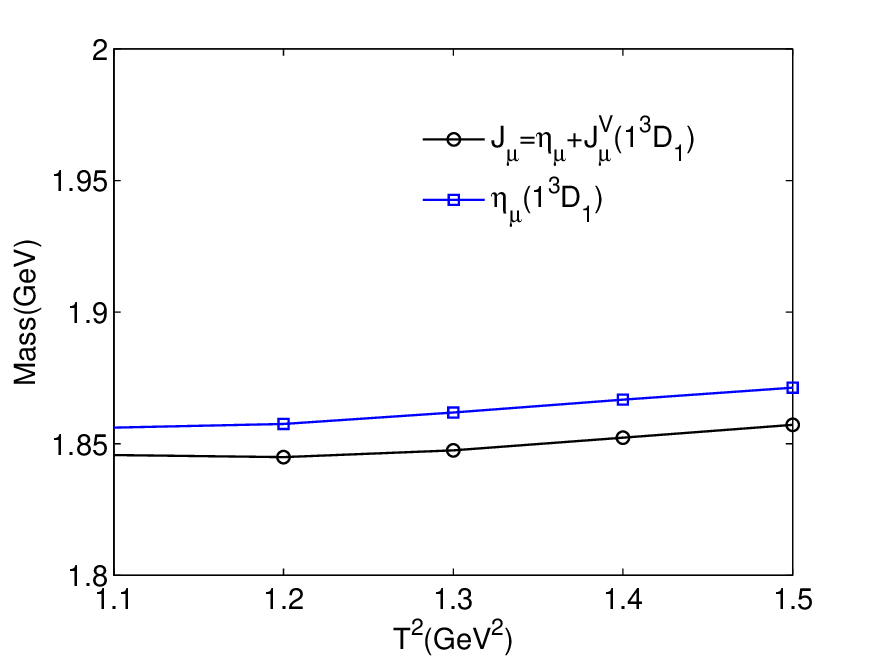}
\caption{The mass of $\phi$($1^{3}D_{1}$) state for different values
of the Borel parameter $T^{2}$
, where the currents
$J_{\mu}$ and $\eta_{\mu}$ are considered respectively..\label{your label}}
\end{minipage}
\hfill
\begin{minipage}[t]{0.45\linewidth}
\centering
\includegraphics[height=5cm,width=7cm]{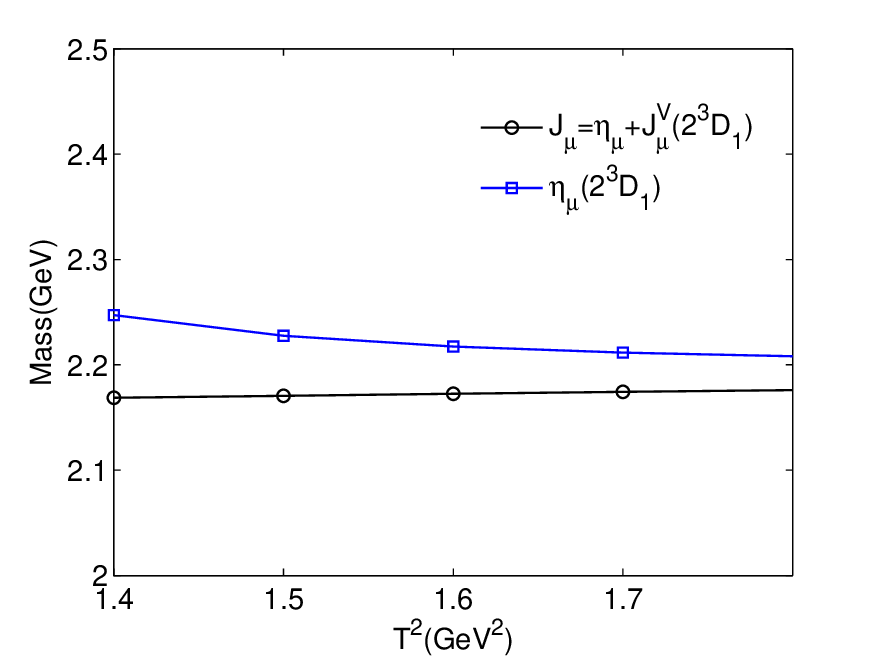}
\caption{The mass of $\phi$($2^{3}D_{1}$) state for different values
of the Borel parameter $T^{2}$
, where the currents
$J_{\mu}$ and $\eta_{\mu}$ are considered respectively.\label{your label}}
\end{minipage}
\end{figure}
\begin{figure}[h]
\begin{minipage}[t]{0.45\linewidth}
\centering
\includegraphics[height=5cm,width=7cm]{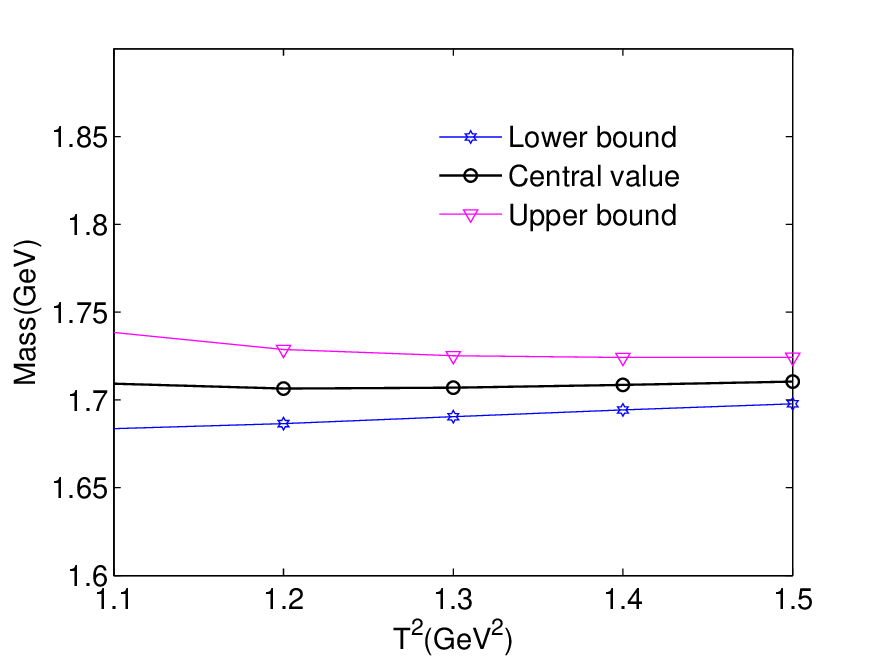}
\caption{The mass of $\rho$($1^{3}D_{1}$) state with variations of the Borel parameter $T^{2}$.\label{your label}}
\end{minipage}
\hfill
\begin{minipage}[t]{0.45\linewidth}
\centering
\includegraphics[height=5cm,width=7cm]{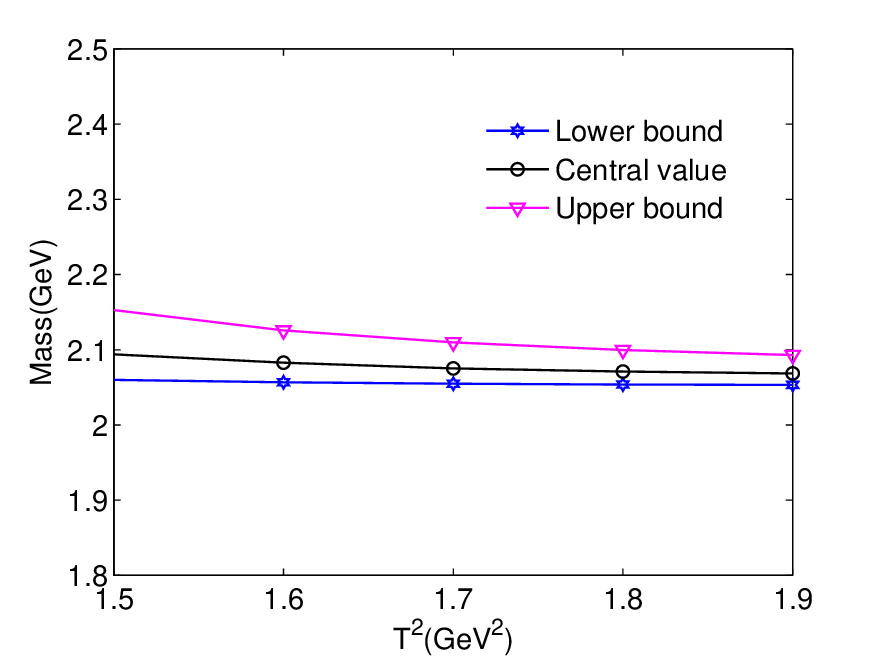}
\caption{The mass of $\rho$($2^{3}D_{1}$) state with variations of the Borel parameter $T^{2}$.\label{your label}}
\end{minipage}
\end{figure}
\begin{figure}[h]
\begin{minipage}[t]{0.45\linewidth}
\centering
\includegraphics[height=5cm,width=7cm]{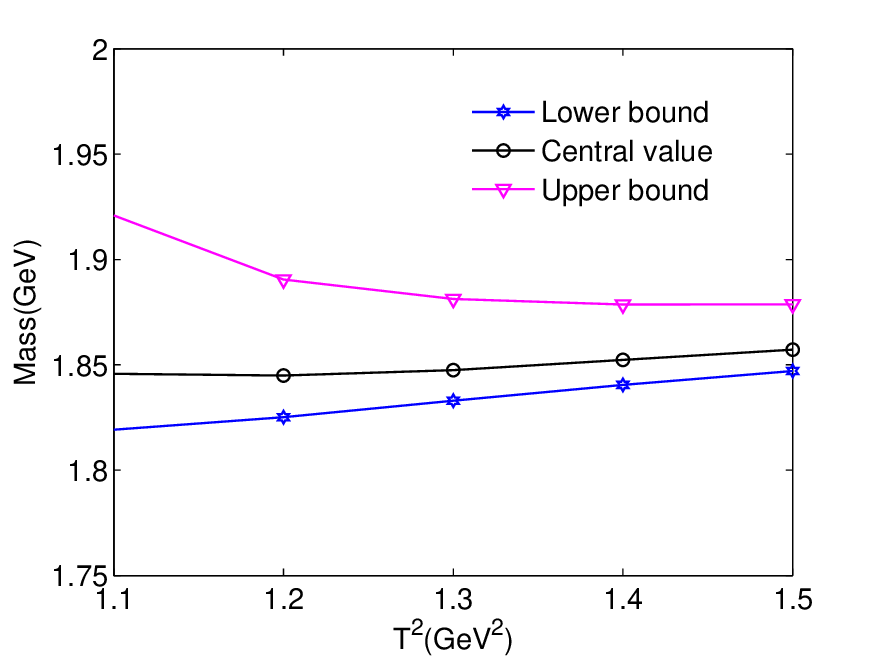}
\caption{The mass of $\phi$($1^{3}D_{1}$) state with variations of the Borel parameter $T^{2}$.\label{your label}}
\end{minipage}
\hfill
\begin{minipage}[t]{0.45\linewidth}
\centering
\includegraphics[height=5cm,width=7cm]{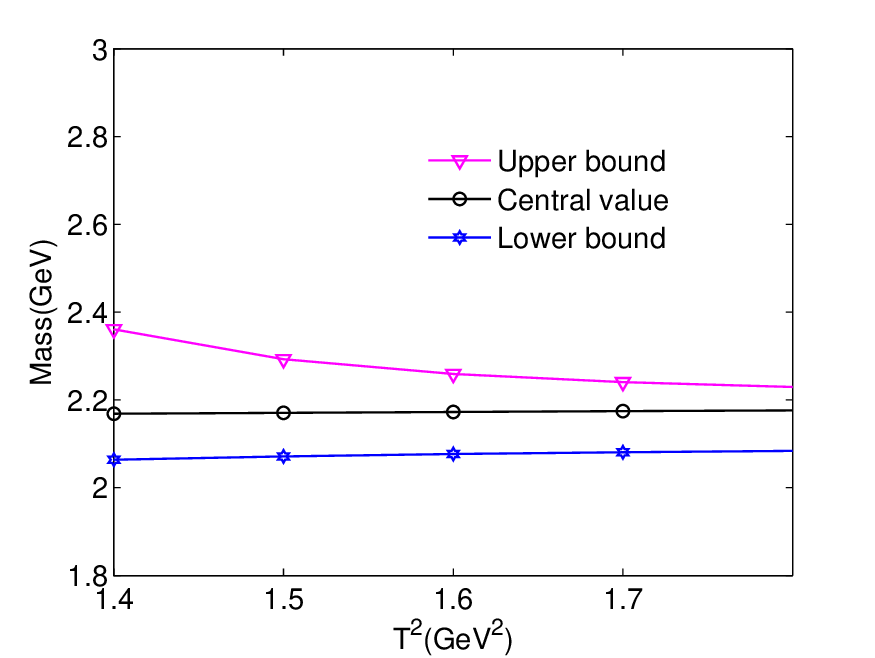}
\caption{The mass of $\phi$($2^{3}D_{1}$) state with variations of the Borel parameter $T^{2}$.\label{your label}}
\end{minipage}
\end{figure}
\begin{figure}[h]
\begin{minipage}[t]{0.45\linewidth}
\centering
\includegraphics[height=5cm,width=7cm]{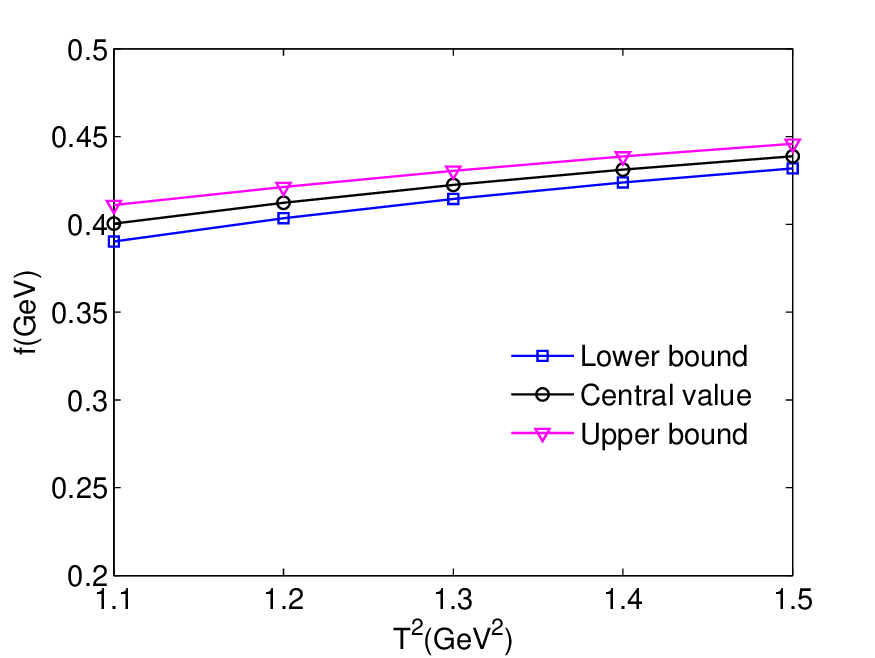}
\caption{The decay constant $\rho$($1^{3}D_{1}$) with variations of the Borel parameter $T^{2}$.\label{your label}}
\end{minipage}
\hfill
\begin{minipage}[t]{0.45\linewidth}
\centering
\includegraphics[height=5cm,width=7cm]{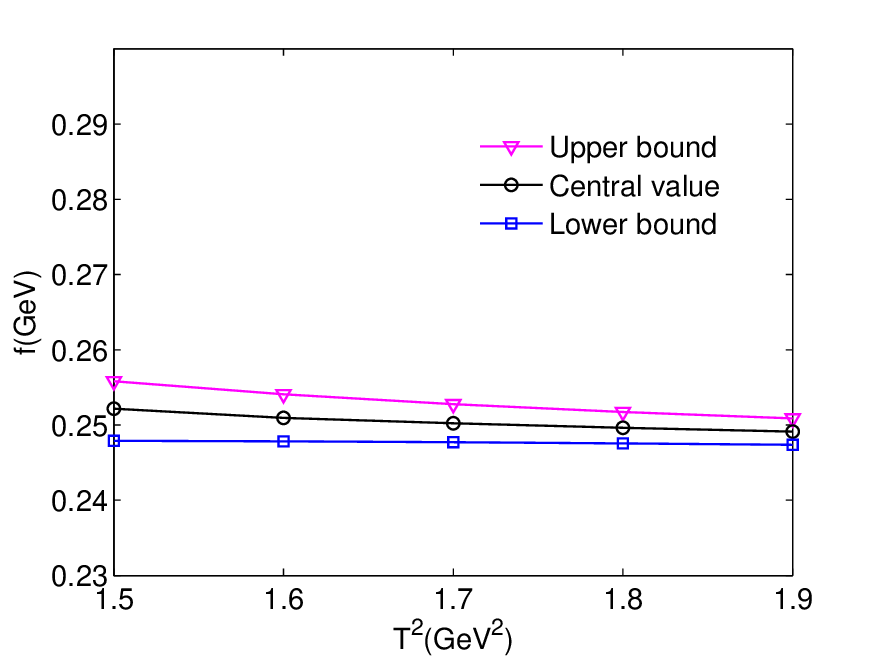}
\caption{The decay constant $\rho$($2^{3}D_{1}$) with variations of the Borel parameter $T^{2}$.\label{your label}}
\end{minipage}
\end{figure}
\begin{figure}[h]
\begin{minipage}[t]{0.45\linewidth}
\centering
\includegraphics[height=5cm,width=7cm]{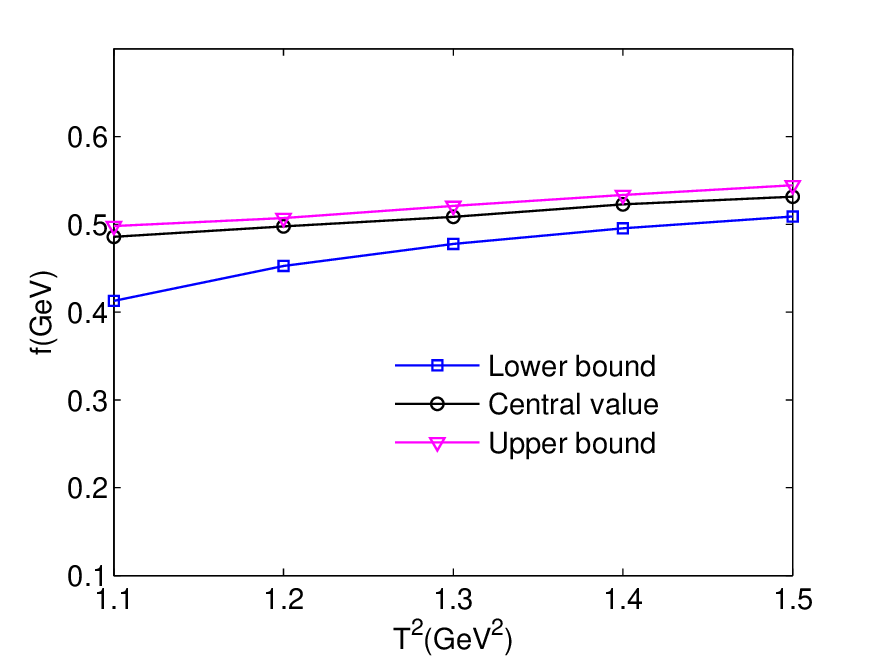}
\caption{The decay constant $\phi$($1^{3}D_{1}$) with variations of the Borel parameter $T^{2}$.\label{your label}}
\end{minipage}
\hfill
\begin{minipage}[t]{0.45\linewidth}
\centering
\includegraphics[height=5cm,width=7cm]{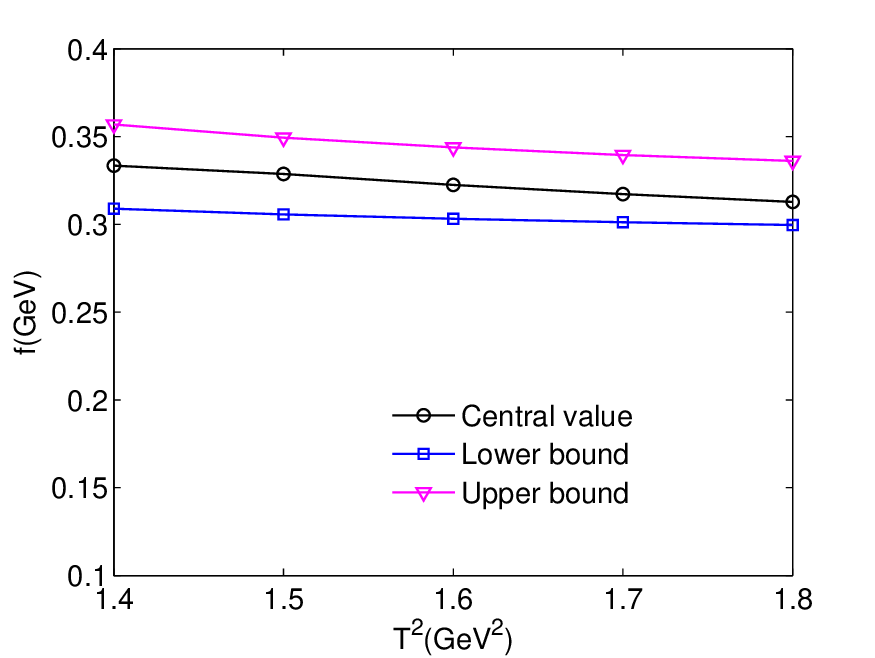}
\caption{The decay constant $\phi$($2^{3}D_{1}$) with variations of the Borel parameter $T^{2}$.\label{your label}}
\end{minipage}
\end{figure}

\begin{table*}[htbp]
\begin{ruledtabular}\caption{The masses of ground state and the first radial excited state of vector D-wave $\rho$ and $\phi$ mesons. All values in units of GeV.}
\begin{tabular}{c| c c c c c c c}
  State& \ This work  & \ GI &\ Ebert  &\ MGI &\ Bug &\ ERT   &\ Exp\\
 \hline
$\phi(1^{3}D_{1})$ &\  $1.851\pm0.095$      &\  $1.876$\cite{GI}   &\ $ 1.845$\cite{Ebert}   &\  $1.869$\cite{MGI} &\ $-$ &\ $1.835$\cite{ERT} &\  $-$ \\
$\phi(2^{3}D_{1})$ &\  $2.176\pm0.075$      &\  $2.337$\cite{GI}   &\ $ 2.258$\cite{Ebert}  & \ $2.276$\cite{MGI}  &\ $-$ &\ $2.185$\cite{ERT} &\ $2.160\pm0.080$\cite{PDG2020}  \\
$\rho(1^{3}D_{1})$&\  $1.708\pm0.040$     &\  $1.660$\cite{GI},$1.664$\cite{DeMin}   &\ $1.570$\cite{Ebert}  & \ $1.646$\cite{DeMin}   &\ $1.700$\cite{Bugg} &\ $-$ &\ $1.720\pm0.020$\cite{PDG2020} \\
$\rho(2^{3}D_{1})$ &\  $2.060\pm0.044$     &\  $2.150$\cite{GI},$2.153$\cite{DeMin}   &\ $1.900$\cite{Ebert} &\  $2.048$\cite{DeMin}  &\ $2.000$\cite{Bugg} &\ $-$ &\ $2.000\pm0.030$\cite{rou21,rou22,rou23,rou24}  \\
\end{tabular}
\end{ruledtabular}
\end{table*}

Finally, we can easily obtain the values of the masses for $\rho$($1^{3}D_{1}$), $\rho$($2^{3}D_{1}$), $\phi$($1^{3}D_{1}$) and $\phi$($2^{3}D_{1}$) states, which are shown in Figs.13-16 and Table II. After taking into account the uncertainties of the input parameters, the uncertainties of the masses are also presented in Figs.13-16 which are marked as the Upper bound and Lower bound. Using these predicted masses as input parameters, we also obtain the decay constants from Eq.(22) and Eq.(28), which are presented in Figs.17-20 and Eq.(37). These estimated decay constants can be used to study the strong decay properties involving $\rho$($1^{3}D_{1}$), $\rho$($2^{3}D_{1}$), $\phi$($1^{3}D_{1}$) and $\phi$($2^{3}D_{1}$) with three-point QCD sum rules or light-core QCD sum rules.

There is no doubt that $\rho$($1700$) is a good candidate of $1^{3}D_{1}$ $\rho$ meson\cite{PDG2020}. On this base, scientists estimated the mass of $2^{3}D_{1}$ $\rho$ meson to be $\sim2.000$ GeV\cite{LPHe,Bugg}. From Table II, we can see that the predicted masses in this work for $\rho$($1^{3}D_{1}$) and $\rho$($2^{3}D_{1}$) are  $1.708\pm0.040$ GeV and $2.060\pm0.044$ GeV. Recently, an resonance was observed in the $\omega\pi^{0}$ cross section by BESIII collaboration which was denoted as $Y$($2040$)\cite{BESIII}. This structure was predicted to be the $2^{3}D_{1}$ $\rho$ meson by its mass ($2034\pm14\pm9$)MeV$/c^{2}$ and decay width ($234\pm11\pm$)MeV. Our predicted mass about this state is consistent well with the experimental data, which means it is reasonable to assign $Y$($2040$) as the $2^{3}D_{1}$ $\rho$ meson.

 Although the ground state of D-wave vector $s\overline{s}$ meson is still not observed in experiment, recently Particle Data Group assigned $\phi$($2170$) to be the $2^{3}D_{1}$ $\phi$ meson with a mass of $2.16$ GeV\cite{PDG2020}. In addition, BESIII collaboration reported the mass of $\phi$($2170$) to be ($2179\pm21\pm3$)MeV/$c^{2}$ in the decay process $\omega\eta$\cite{BESIII} From Table II, we can see that the predicted masses of $2^{3}D_{1}$ $\phi$ meson with different theoretical methods are not agreement well with each other. Our predicted mass for $2^{3}D_{1}$ $\phi$ meson is $2.176\pm0.075$ GeV which agrees well with experimental data. In addition, we predict the mass of $1^{3}D_{1}$ $\phi$ meson to be $1.851\pm0.095$ GeV which is roughly compatible with those of other collaborations. This prediction is helpful in the future to search for this missing ground state in experiment.
\begin{eqnarray}
\notag\
f_{\rho(1^{3}D_{1})}=0.421\pm0.019GeV \\
\notag\
f_{\rho(2^{3}D_{1})}=0.250\pm0.012GeV \\
f_{\phi(1^{3}D_{1})}=0.509\pm0.034GeV \\
\notag\
f_{\phi(2^{3}D_{1})}=0.345\pm0.021GeV
\end{eqnarray}

\begin{large}
\textbf{4 Conclusions}
\end{large}

In the past decades, more and more $\rho$ and $\phi$ states have been observed in experiments. How to categorize these states into the meson family is a interesting topic, which can improve our knowledge of light meson spectrum. In this work, we study the masses and decay constants of the ground state and the first radially excited state of D-wave vector $\rho$ and $\phi$ mesons with the QCD sum rules.
Our calculation successfully reproduce the experimental data of $1^{3}D_{1}$ $\rho$ meson and $2^{3}D_{1}$ $\phi$ meson, which indicates our analysis is reliable. We predict the mass of $2^{3}D_{1}$ $\rho$ meson to be $2.060\pm0.044$ GeV, which is accordance with the mass of recently observed $Y$($2040$) state\cite{BESIII}. This result supports assigning $Y$($2040$) resonance to be the $2^{3}D_{1}$ state. We also predict the mass of $1^{3}D_{1}$ $\phi$ meson to be $1.851\pm0.095$ GeV. This result are roughly compatible with the values of other collaborations.
Using the obtained masses as input parameters in QCDSR I and QCDSR II, we finally predict the decay constants for these meson states. The theoretical analysis in this work is not only helpful to confirm the underlying properties of these light mesons, but also serve further experimental investigation.


\begin{large}
\textbf{Acknowledgment}
\end{large}

This work has been supported by the Fundamental Research Funds for the Central Universities, Grant Number $2016MS133$, Natural Science Foundation of HeBei Province, Grant Number $A2018502124$.

\end{document}